\documentclass[aps,prb,twocolumn]{revtex4}

\usepackage{graphicx,color}
\usepackage{amsmath}
\usepackage{dcolumn}

\begin{document}
\title{Lyman spectra of holes bound to Cu, Ag, and Au acceptors in ZnTe and CdTe}
\author{Gang Chen}
\altaffiliation[Current address: ]{Department of Physics, University of California San Diego, La Jolla, California 92093, USA}
\author{I. Miotkowski}
\author{A. K. Ramdas}
\affiliation{Department of Physics, Purdue University, West Lafayette, Indiana 47907, USA}
\date{\today}
\begin{abstract}
The group IB impurities (Cu, Ag, and Au) incorporated into II-VI zinc blende hosts of ZnTe and CdTe exhibit well resolved excitation lines followed by a photoionization continuum in their infrared absorption spectra. They are associated with transitions from a \textquotedblleft $1s$-like\textquotedblright ~ground state to various \textquotedblleft $p$-like\textquotedblright ~excited state characteristic of a hole bound to a Coulomb center. Their spacing agree well with those predicted in the effective mass theory for single acceptors as expected for group IB elements substitutionally replacing the group IIB cations of the host. The occurrence of the simultaneous excitation of the Lyman transitions in combination with the zone center longitudinal optical phonon and hence lying in the photoionization continuum and displaying Fano-like asymmetries are features described and interpreted. 
\end{abstract}
\maketitle

\section{Introduction}
As is well known, impurities which trap electrons (\textit{donors}) and holes (\textit{acceptors}) are essential for exploiting semiconductors in solid state electronics and optical sensors. In the context of the physics of semiconductors, donors and acceptors are noteworthy for the manner in which the nature of the bound and continuum states of the charge carriers bound to them arise, on the one hand, and the band extrema with which they are associated, on the other. Bound and continuum states of donors and acceptors have been studied intensively exploiting absorption, emission, and photoconductivity spectra.

\begin{figure}
\resizebox{3.4in}{!}{\includegraphics{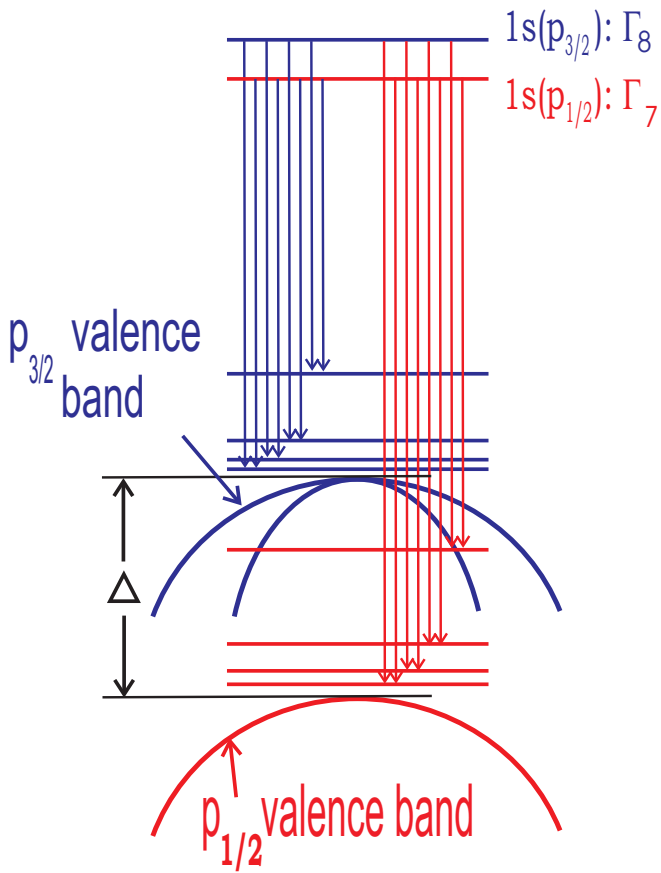}}
\caption{\label{FIG1}Energy levels of a hole bound to a substitutional group IB accepter in a zincblende II-VI semiconductor are shown schematically. The diagram also shows the Lyman transitions from the ground state associated with the $p_{3/2}$ valence band as well as those associated with the $p_{1/2}$ valence band. $\Delta$ is the energy difference between the maxima of the two bands.}
\end{figure}

The focus in the present work is on acceptors in II-VI semiconductors. Consider a substitutional group IB element replacing a group IIB cation [e.g., Cu$^{+}$ replacing Zn$^{2+}$ in ZnTe (i.e., Cu$_{\mathrm{Zn}}$)] in a II-VI semiconductor. It borrows an electron from the valence band and creates a hole in order to complete the tetrahedral bonds with the nearest neighbor group VI anions of the host. At sufficiently low temperatures, the hole is bound to the impurity ion via a screened Coulomb attraction. The bound states of the hole are shown schematically in Fig.~\ref{FIG1}. The binding energy of this solid state analog of the hydrogen atom is very small compared to that of atomic hydrogen as a consequence of the large dielectric constant of the host semiconductor and the small effective mass of the bound hole. In the effective mass theory (EMT)~\cite{Kohn}, including spin-orbit effects, these states are associated with the $p_{3/2}(\Gamma_{8})$ and $p_{1/2}(\Gamma_{7})$ valence band maxima located at the zone center. The wave function of the bound states are products of hydrogenic envelope functions and Bloch functions characteristic of the valence band maxima. Consistent with the $T_d$ symmetry of the substitutional acceptor, the bound states are characterized by $\Gamma_6$, $\Gamma_7$, or $\Gamma_8$ symmetry. The $1s(p_{3/2})$ and $1s(p_{1/2})$ ground states have $\Gamma_8$ and $\Gamma_7$ symmetry, respectively. The binding energies of the $s$-like and $p$-like states are determined by the effective mass parameters, or equivalently by the Luttinger parameters~\cite{Luttinger}, and the dielectric constant of the host. In addition, the bound states might experience shifts originating in the departures from EMT due to the so called \textquotedblleft central cell\textquotedblright ~corrections~\cite{Ramdas,Pantelides}. At low temperatures, the electric-dipole allowed transitions from the $1s(\Gamma_8)$ to the excited $p$-states, i.e., the Lyman lines can be observed.

The II-VI semiconductors usually prefer one type over the other, for example, ZnTe is known to be p-type~\cite{Bittebierre}. Apart from the non-stoichiometric nature of these crystals which results in vacancies and interstitials, there are some major residual impurities dominating the crystal properties. It is of particular interest to obtain the detailed information on the electronic behavior of these impurities. 

It has been established that Cu and Ag are two common residual impurities in ZnTe and CdTe, which control the electrical and optical properties of these materials~\cite{Dean,Chamonal}. Previous studies were performed on samples in which the impurities were introduced after growth by thermal diffusion~\cite{Saminadayar,Molva}. While these impurities appear in as-grown ZnTe and CdTe crystals, studies on crystals with impurities introduced during growth are needed. 

The observation and discussion of the Lyman spectra of Cu, Ag, and Au acceptors in ZnTe and CdTe along with their phonon replicas is the subject of the present study. In the course of the investigation, a fascinating phenomenon called \textquotedblleft Fano resonance\textquotedblright ~was observed in their Lyman spectra. They are discussed in terms of parameters associated with the coupling of the discrete bound states of the acceptor as LO-phonon assisted replicas overlapping the continuum states.

\section{Experimental}
Single crystals of ZnTe and CdTe were grown by vertical Bridgman technique using a two-zone furnace with a linear temperature gradient of 5 $^{\circ}$C/cm in the growth zone.~\cite{Miotkowski} Cu, Ag and Au impurities were introduced by adding corresponding ingredient into the starting materials.

The absorption spectra were recorded using an ultrahigh resolution BOMEM DA.3 Fourier transform infrared (FTIR) spectrometer~\cite{BOMEM} capable of an ultimate unapodized resolution of 0.0026 cm$^{-1}$. A HgCdTe-infrared detector cooled with liquid nitrogen were employed. A Janis 10DT Supervaritemp~\cite{Janis} optical cryostat, with polypropylene windows along one axis and wedged ZnSe inner and CsI outer windows along the other axis, allowed measurements in the range of 1.8-300 K. The specimens were prepared for infrared absorption measurements by suitably polishing two approximately plane parallel surfaces, with a small wedge introduced to avoid channelling in the spectrum.

\section{Experimental results and discussion}
\label{AcceptorsExp}
\subsection{Cu, Ag, and Au acceptors in ZnTe}
\begin{figure*}
\resizebox{4.5in}{!}{\includegraphics{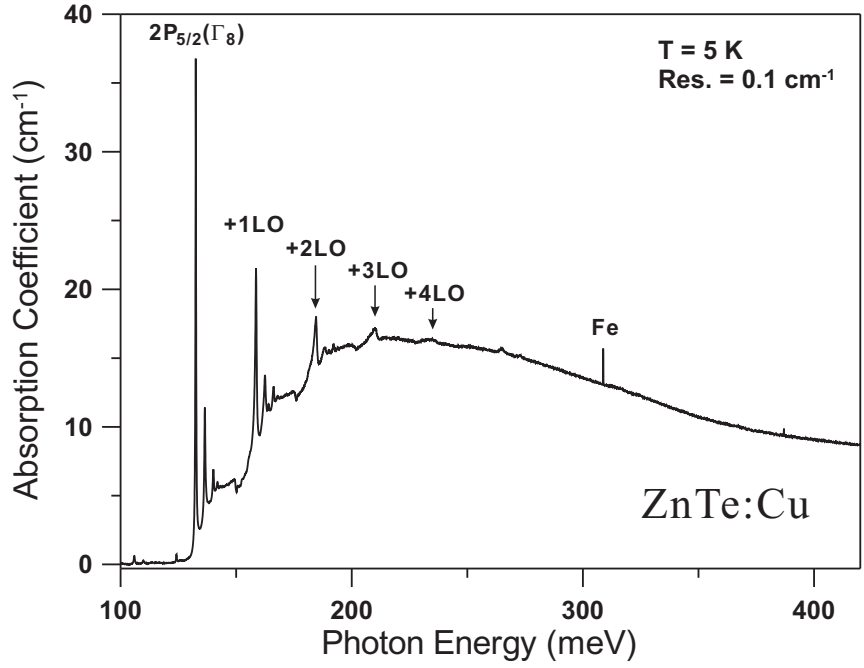}}
\caption{\label{FIG2}The Lyman spectrum of Cu acceptors in ZnTe along with their 1LO, 2LO, 3LO and 4LO replicas.}
\end{figure*}

The absorption spectrum of ZnTe doped with Cu recorded at temperature, $T = 5$ K, is displayed in Fig.~\ref{FIG2}. The sharp excitation lines of the Cu acceptors in the range 120-150 meV are due to the Lyman transitions from the $1s_{3/2} (\Gamma_8)$ ground state to the $np$ ($n = 2, 3, \dots$) excited states. They are followed by a photoionization continuum with its onset $\sim 149$ meV, representing transitions from the ground state of Cu acceptor to the \textit{p}$_{3/2}$ valence band of ZnTe. The lines, superposed on the continuum, are interpreted as the Lyman excitations in combination with one, two, three, and four quanta of the zone center longitudinal optical phonon (LO) of the ZnTe host, starting at 158.6, 184.52, 209.98, and 234.52 meV for the $1s_{3/2}(\Gamma_8) \rightarrow 2p_{5/2}(\Gamma_8)$ transition, respectively. The polar character of ZnTe clearly singles out the special role of the LO phonon and its overtones, the microscopic mechanism for the coupling being the Fr\"{o}hlich interaction~\cite{Yu2}. The positions of the observed excitation lines along with the values reported by Saminadayar \textit{et al.}~\cite{Saminadayar} are listed in Table~\ref{Tab:ZnTe1}. The absorption due to the photoionization continuum increases with photon energy initially, reaching a maximum $\sim 200$ meV and followed by a monotonic decrease. The physical process underlying the photoionization involves a product of the oscillator strength for the transitions from the $1s$ initial state to final states in the heavy hole and light hole $p_{3/2}$ valence bands and their density of states.

\begin{figure*}
\resizebox{4.5in}{!}{\includegraphics{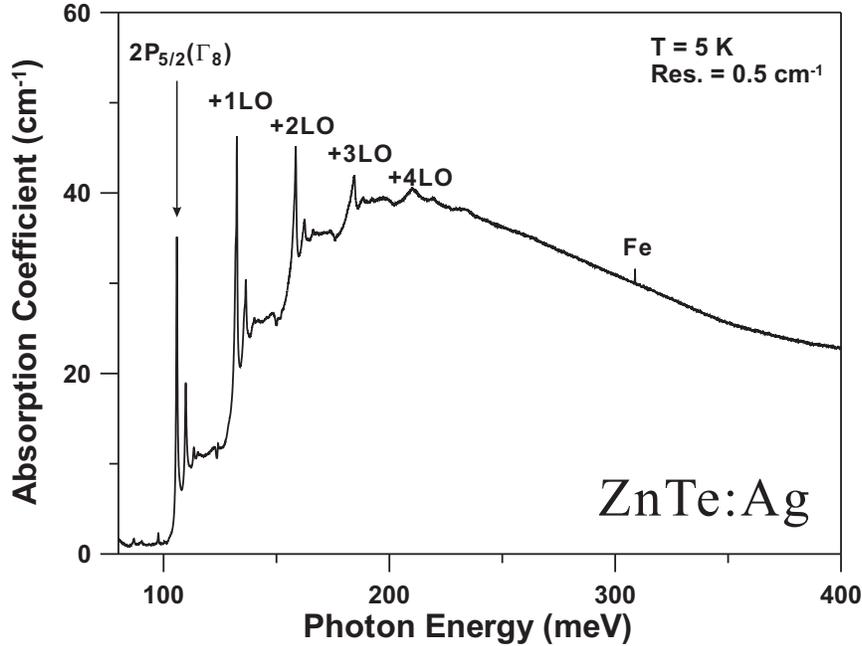}}
\caption{\label{FIG3}The Lyman spectrum of Ag acceptors in ZnTe along with their 1LO, 2LO, 3LO and 4LO replicas.}
\end{figure*}

In a similar fashion, Fig.~\ref{FIG3} shows the absorption spectrum of ZnTe doped with Ag recorded at $T = 5$ K. The Lyman spectrum arising from the $1s \rightarrow np$ transitions for the Ag acceptors are observed in the range 90-120 meV. The photoionization continuum starting at $\sim 122$ meV increases to a maxima at $\sim 200$ meV and then slowly decreases. As in the case of the Cu acceptors, the phonon replicas are observed up to the fourth overtone of the zone center LO phonon with onsets at 131.99, 158.08, 184.34, and 209.52 meV for the $1s_{3/2}(\Gamma_8) \rightarrow 2p_{5/2}(\Gamma_8)$ transition. The positions of Lyman excitations and their phonon replicas are listed in Table~\ref{Tab:ZnTe1}. 
 
\begin{table*}
\caption{\label{Tab:ZnTe1}Observed transition energies from $1s_{3/2}(\Gamma_8)$ ground state to $np$ excited states of Cu, Ag and Au acceptors in ZnTe (in meV) compared with the results of Ref.~\onlinecite{Saminadayar}.}
\begin{ruledtabular}
\begin{tabular}{cccccccc} 
Final State & Cu & Cu\footnotemark[1] & Ag & Ag\footnotemark[1] & Au & Au\footnotemark[1] \\ 
\hline
$2p_{3/2}(\Gamma_8)$     & 124.18 &       & 97.69  & 100   &        &       \\
$2p_{5/2}(\Gamma_8)$     & 132.52 & 134   & 105.88 & 107.5 & 251.67 & 255.2 \\
$2p_{5/2}(\Gamma_7)$     & 136.46 & 138.3 & 109.82 & 111.5 & 255.76 & 259   \\
$3p_{5/2}(\Gamma_8)$     & 140.15 &       & 113.39 &       &        &       \\
$3p_{5/2}(\Gamma_7)$     & 141.92 &       & 115.21 &       &        &       \\
$4p_{5/2}(\Gamma_8)$     & 143.28 &       &        &       &        &       \\
1LO+$2p_{5/2}(\Gamma_8)$ & 158.60 & 160.3 & 131.99 & 134.2 & 277.41 & 281.1 \\
1LO+$2p_{5/2}(\Gamma_7)$ & 162.45 & 164.6 & 135.93 & 137.4 & 281.79 & 285   \\
1LO+$3p_{5/2}(\Gamma_8)$ & 166.18 &       & 139.58 &       &        &       \\
2LO+$2p_{5/2}(\Gamma_8)$ & 184.52 & 186.2 & 158.08 & 159.6 & 302.50 & 307.7 \\
2LO+$2p_{5/2}(\Gamma_7)$ & 188.34 &       & 162.31 &       &        &       \\
2LO+$3p_{5/2}(\Gamma_8)$ & 192.13 &       & 166.11 &       &        &       \\
3LO+$2p_{5/2}(\Gamma_8)$ & 209.98 & 212.2 & 184.34 & 185   & 328.54 & 333.9 \\
3LO+$2p_{5/2}(\Gamma_7)$ & 218.12 &       & 188.29 &       &        &       \\
4LO+$2p_{5/2}(\Gamma_8)$ & 234.52 &       & 209.89 &       &        &       \\
\end{tabular}
\end{ruledtabular}
\footnotetext[1]{See Ref. \onlinecite{Saminadayar}}
\end{table*}

As is well known the effective mass theory~\cite{Ramdas} is successful in predicting the binding energies of the excited $p$-like states whose wave functions have negligible amplitudes at the Coulomb center; in contrast, the wave function of the $1s$-like ground state has a large amplitude at the impurity site. As a consequence, the ground state is sensitive to the chemical nature of the acceptor impurity and its binding energy $E_{I}$ experiences a departure from the effective mass value by an amount characteristic for a given impurity. The Lyman spectra for different group IB impurities thus are expected to be identical as far as the spacings of the corresponding lines and their relative intensities are concerned, but shifted in energy with respect to one another due to this chemical shift of the $s$-like ground state.

\begin{figure*}
\resizebox{5in}{!}{\includegraphics{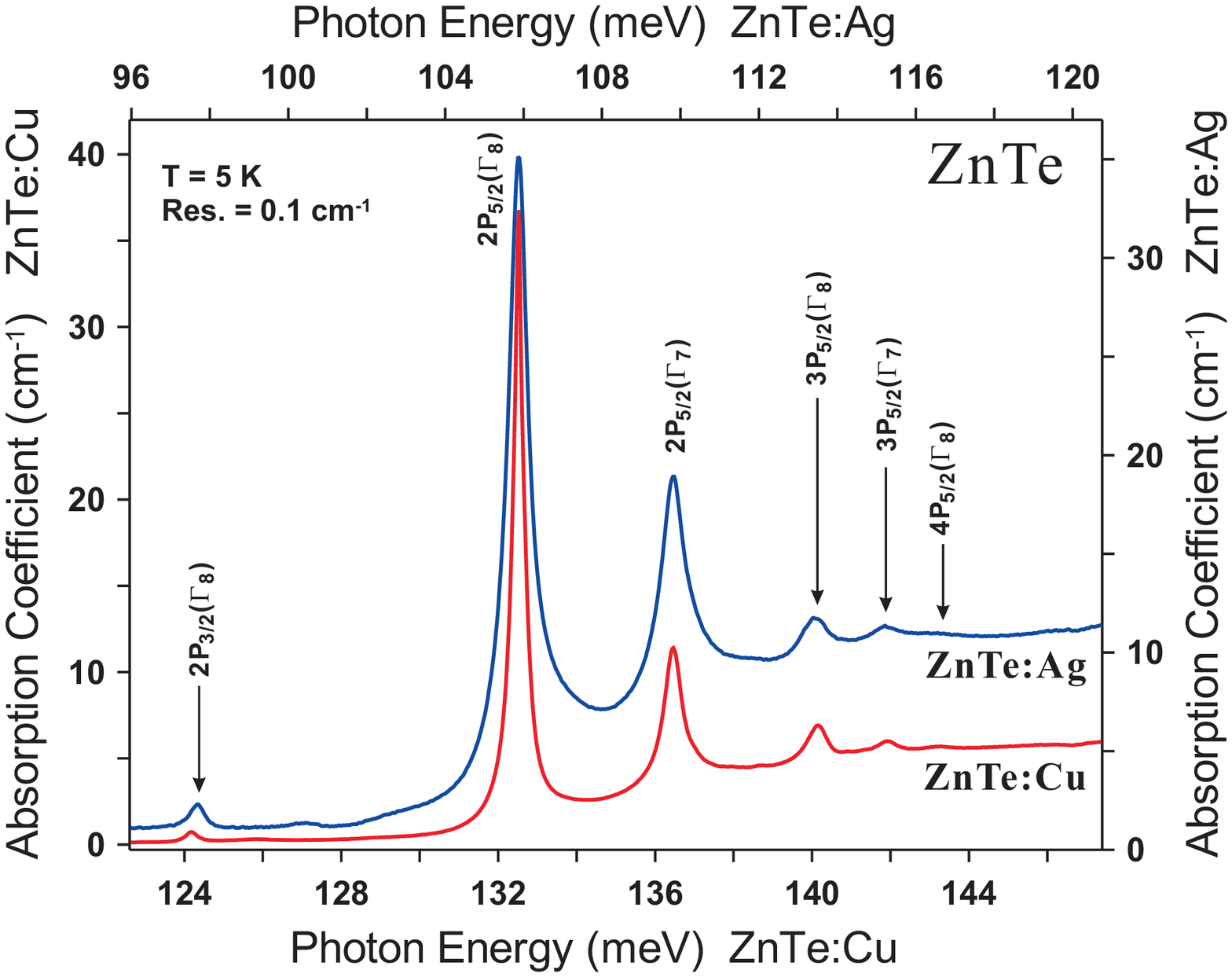}}
\caption{\label{FIG4}The Lyman spectra of ZnTe:Cu and ZnTe:Ag compared with the $2p_{3/2}(\Gamma_{8})$ transition in each brought into coincidence. The energy scale for the two spectra being the same, the identity of the energy spacings is made evident in the figure.}
\end{figure*}

\begin{table*}
\caption{\label{Tab:ZnTe2}Binding energies (in meV) of the final states of the observed transitions of Cu, Ag, and Au in ZnTe.}
\begin{ruledtabular}
\begin{tabular}{cccccccc}
Final State & Cu & Cu\footnotemark[1] & Ag & Ag\footnotemark[1] & Au & Au\footnotemark[1] & Theory\footnotemark[2]\\ 
\hline
$1s_{3/2}(\Gamma_8)$ & 148.82 & 150.3 & 122.18 & 123.8 & 267.97 & 271.6 & 62.3 \\
$2p_{3/2}(\Gamma_8)$ & 24.64  &       & 24.49  & 23.8  &        &       & 23.6 \\
$2p_{5/2}(\Gamma_8)$ & 16.3   & 16.3  & 16.3   & 16.3  & 16.3   & 16.4  & 16.3 \\
$2p_{5/2}(\Gamma_7)$ & 12.36  & 12.0  & 12.36  & 12.3  & 12.21  & 12.6  & 13.05 \\
$3p_{5/2}(\Gamma_8)$ & 8.67   &       & 8.79   &       &        &       & \\
$3p_{5/2}(\Gamma_7)$ & 6.9    &       & 6.97   &       &        &       & \\
$4p_{5/2}(\Gamma_8)$ & 5.54   &       &        &       &        &       & \\
\end{tabular}
\end{ruledtabular}
\footnotetext[1]{See Ref. \onlinecite{Saminadayar}}
\footnotetext[2]{See Ref. \onlinecite{Herbert}}
\end{table*}

The Lyman spectra of Cu and Ag acceptors in ZnTe are displayed in Fig.~\ref{FIG4} on the same scale with their most prominent line at 2\textit{p}$_{5/2}(\Gamma_{8})$ brought into coincidence. The identity of the spacings of the spectra for the two acceptor impurities is immediately apparent. The spacings, in turn, are close to those predicted using the perturbation theory first employed by Baldereschi and Lipari~\cite{Baldereschi,Herbert} for single acceptors in ZnTe and thus provide the justification for labeling the excitation lines on the basis of the final state of the observed transitions. With this interpretation, one can add the calculated binding energy of a given transition to its experimental transition energy to deduce $E_{I}$. Since the $p$-like states are only slightly affected by the central cell potential, the ionization energy $E_{I}$ can be deduced by adding the binding energy of $2p_{5/2}(\Gamma_{8})$ to the observed energy of $1s_{3/2}(\Gamma_{8}) \rightarrow 2p_{5/2}(\Gamma_{8})$ transition. In this manner, the ionization energies for Cu, Ag, and Au acceptors in ZnTe are calculated to be 148.82, 122.18, and 267.97 meV, respectively. The binding energies for the final states of the observed transitions of the group IB impurities as single acceptors in ZnTe along with the theoretical values are listed in Table~\ref{Tab:ZnTe2}.

\subsection{Cu and Ag acceptors in CdTe}

\begin{figure*}
\resizebox{4.5in}{!}{\includegraphics{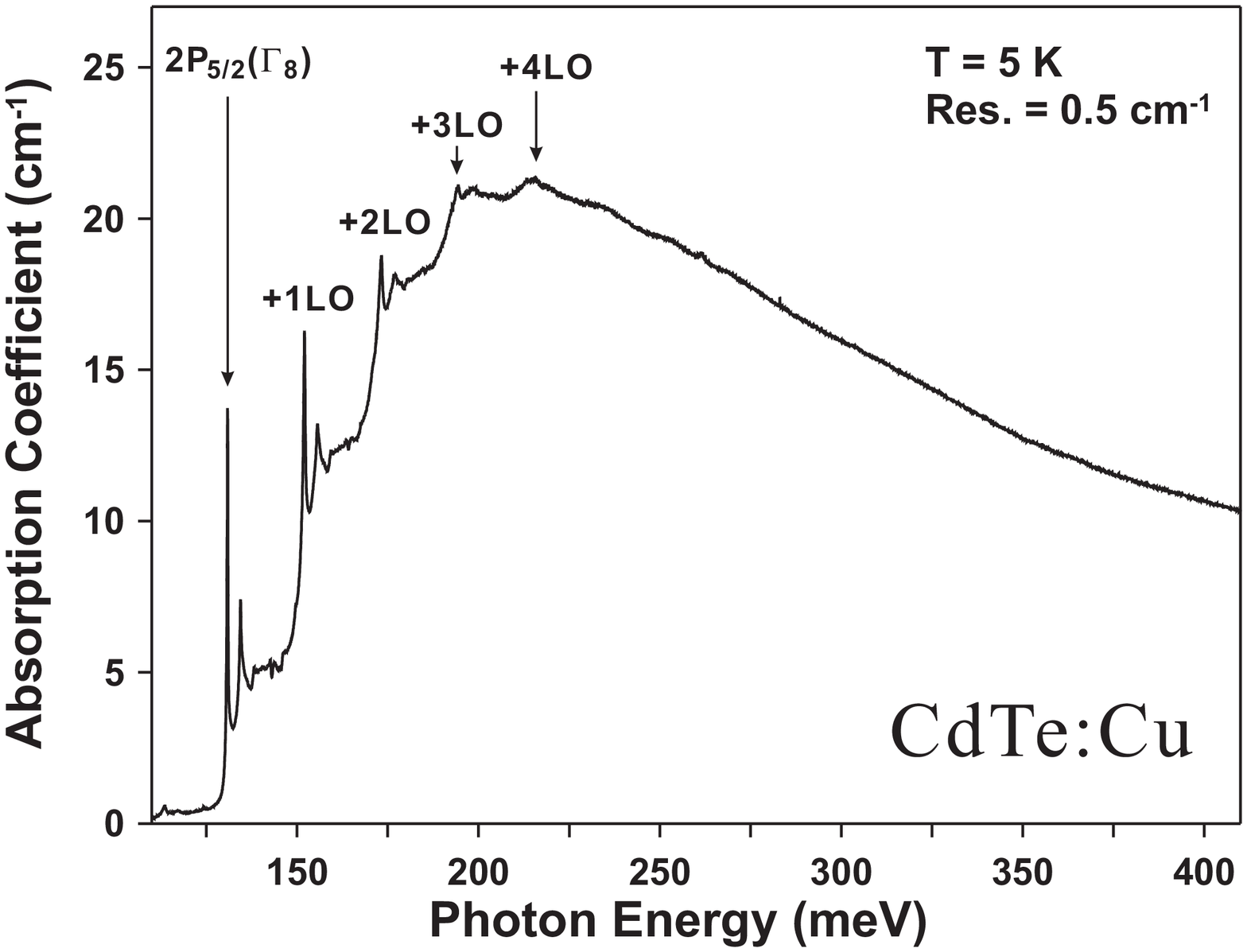}}
\caption{\label{FIG5}The Lyman spectrum of Cu acceptors in CdTe along with their 1LO, 2LO, 3LO and 4LO replicas.}
\end{figure*}

The absorption spectrum of CdTe doped with Cu recorded at $T = 5$ K is displayed in Fig.~\ref{FIG5}. The excitation lines in the range $125-145$ meV are the zero phonon Lyman transitions followed by a photoionization continuum with its onset at $146 \pm 0.5$ meV, representing transitions from the ground state of the Cu acceptor to the $p_{3/2}$ valence band of CdTe. The absorption due to the photoionization continuum increases with photon energy initially, reaching a maximum $\sim 220$ meV and followed by a monotonic decrease. The step-like features are the Lyman excitations in combination with one, two, three, and four overtone of the zone center longitudinal optical phonon of CdTe host with their onsets at 152.17, 173.35, 194.57, and 215.07 meV, respectively. The energies of these observed transitions compared with those reported by Molva \textit{et al.}~\cite{Molva} are listed in Table~\ref{Tab:CdTe1}.

\begin{figure*}
\resizebox{4.5in}{!}{\includegraphics{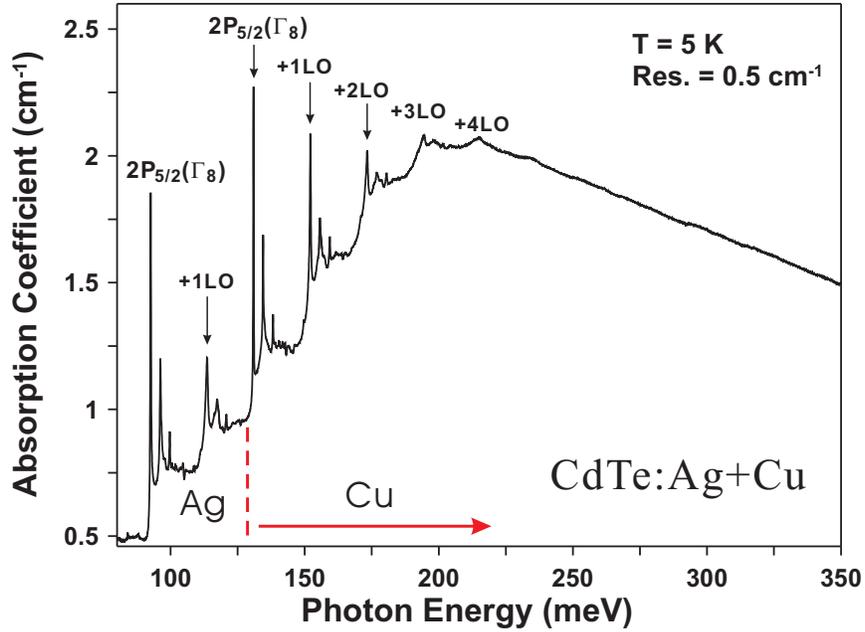}}
\caption{\label{FIG6}The absorption spectrum of a specimen doped with Ag with Cu residual impurity shows the Lyman transitions due to the Ag and Cu acceptors in the ranges of $80 - 130$ and $130 - 225$ meV, respectively. The first phonon replica for the Ag acceptor and up to fourth phonon replicas for the Cu acceptor are labeled in the figure.}
\end{figure*}

\begin{table}
\caption{\label{Tab:CdTe1}Observed transition energies from $1s_{3/2}(\Gamma_8)$ ground state to $np$ excited states of Cu and Ag acceptors in CdTe (in meV) compared with the results of Ref.~\onlinecite{Molva}.}
\begin{ruledtabular}
\begin{tabular}{cccccccc} 
Final State & Cu & Cu\footnotemark[1] & Ag & Ag\footnotemark[1] \\ 
\hline
$2p_{3/2}(\Gamma_8)$     & 124.33 &       &        &        \\
$2p_{5/2}(\Gamma_8)$     & 130.95 & 130.9 & 92.55  & 92.5   \\
$2p_{5/2}(\Gamma_7)$     & 134.56 & 134.6 & 96.17  & 96.2   \\
$3p_{5/2}(\Gamma_8)$     & 138.28 &       & 99.65  &        \\
$3p_{5/2}(\Gamma_7)$     & 140.61 &       &        &        \\
1LO+$2p_{5/2}(\Gamma_8)$ & 152.17 &       & 113.58 &        \\
1LO+$2p_{5/2}(\Gamma_7)$ & 155.75 &       & 117.35 &        \\
1LO+$3p_{5/2}(\Gamma_8)$ & 159.38 &       & 120.75 &        \\
2LO+$2p_{5/2}(\Gamma_8)$ & 173.35 &       &        &        \\
2LO+$2p_{5/2}(\Gamma_7)$ & 177.28 &       &        &        \\
3LO+$2p_{5/2}(\Gamma_8)$ & 194.57 &       &        &        \\
4LO+$2p_{5/2}(\Gamma_8)$ & 215.07 &       &        &        \\
\end{tabular}
\end{ruledtabular}
\footnotetext[1]{See Ref. \onlinecite{Molva}}
\end{table}

The Lyman transitions for Ag acceptors in CdTe from the $1s(p_{3/2})$ ground state to the $np$ states associated with the $p_{3/2}$ valence band are observed in the range $87 - 105$ meV as shown in Fig.~\ref{FIG6}. In this specimen, while Ag was deliberately introduced by doping, Cu appeared as a residual impurity. The Lyman transitions of the Ag acceptor and its first LO phonon replica starting from 113.58 meV are observed in the range $80 - 130$ meV. Beyond this range, the absorption spectrum of the Ag acceptor overlaps with the Lyman series and its up to fourth overtone phonon replicas of the residual Cu acceptor which can be clearly identified in Fig.~\ref{FIG6}. Table~\ref{Tab:CdTe1} lists the observed transition energies for the Ag acceptor and compared them with the results from Molva \textit{et al}~\cite{Molva}.   

\begin{figure*}
\resizebox{5in}{!}{\includegraphics{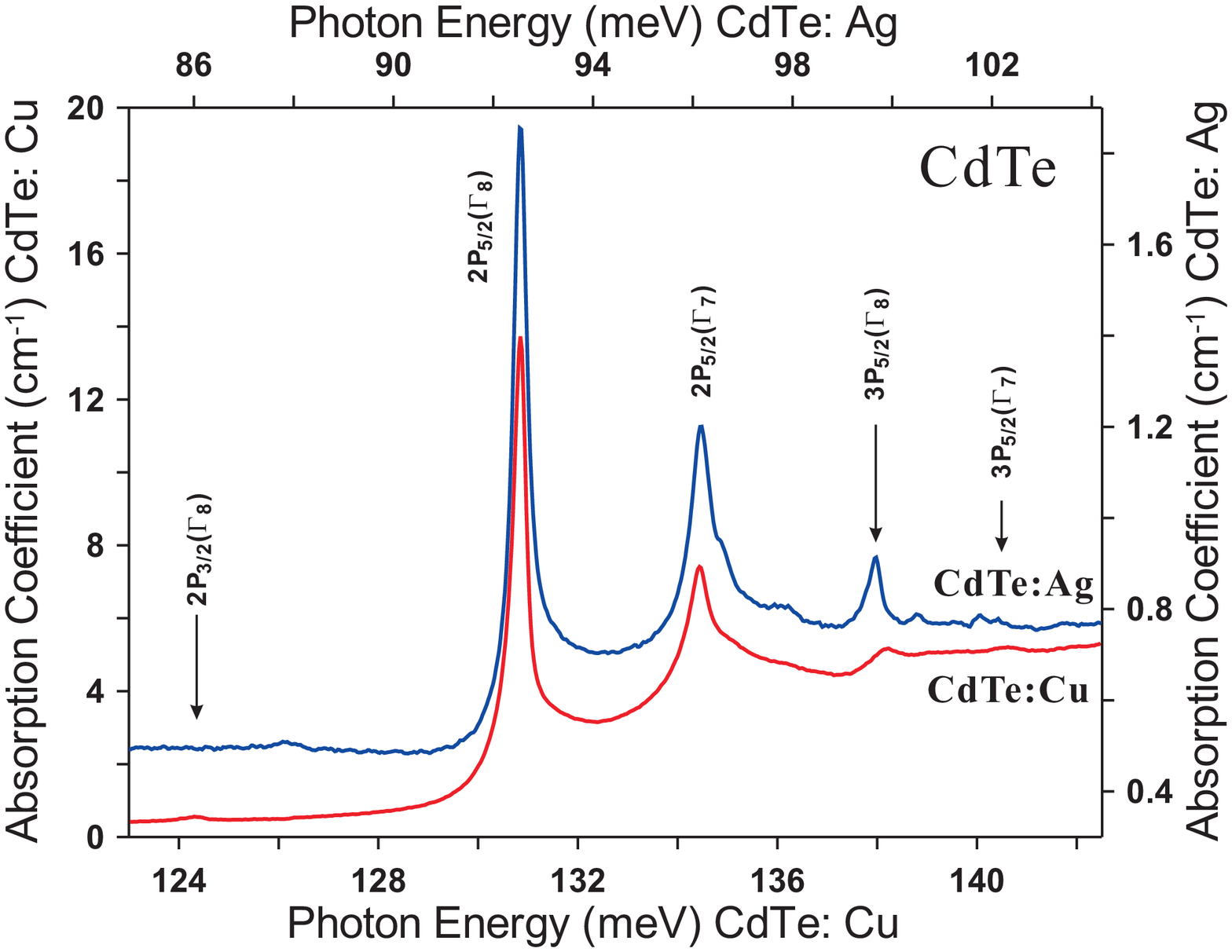}}
\caption{\label{FIG7}The Lyman spectra of CdTe:Cu and CdTe:Ag compared with the $2p_{3/2}(\Gamma_{8})$ transition in each brought into coincidence. The energy scale for the two spectra being the same, the identity of the energy spacings is made evident in the figure.}
\end{figure*}

\begin{table}
\caption{\label{Tab:CdTe2}Binding energies (in meV) of the final states of the observed transitions of Cu and Ag in CdTe.}
\begin{ruledtabular}
\begin{tabular}{cccccc}
Final State & Cu & Cu\footnotemark[1] & Ag & Ag\footnotemark[1] & Theory\footnotemark[2]\\ 
\hline
$1s_{3/2}(\Gamma_8)$ & 146.25 & 146.0 & 107.85 & 107.5  & 56.8 \\
$2p_{3/2}(\Gamma_8)$ & 21.92  & 24.2  &        & 24.2   & 23.8 \\
$2p_{5/2}(\Gamma_8)$ & 15.3   & 15.1  & 15.3   & 15.0   & 15.3 \\
$2p_{5/2}(\Gamma_7)$ & 11.69  & 11.4  & 11.68  & 11.3   & 11.7 \\
$3p_{5/2}(\Gamma_8)$ & 7.97   &       & 8.2    &        & \\
$3p_{5/2}(\Gamma_7)$ & 5.64   &       &        &        & \\
\end{tabular}
\end{ruledtabular}
\footnotetext[1]{See Ref. \onlinecite{Molva}}
\footnotetext[1]{See Ref. \onlinecite{Molva2}}
\end{table}

Similar to the case of ZnTe, one would expect the spacing of the corresponding lines as well as their relative intensities to be same for different group IB impurities in CdTe. As can be seen in Fig.~\ref{FIG7}, the Lyman spectra of Cu and Ag acceptors in CdTe are displayed in the same scale with the most prominent transition [$1s_{3/2}(\Gamma_8) \rightarrow 2p_{5/2}(\Gamma_8)$] brought into coincidence. The transitions to the final states of $2p_{3/2}(\Gamma_8)$, $2p_{5/2}(\Gamma_8)$, $2p_{5/2}(\Gamma_7)$, $3p_{5/2}(\Gamma_8)$, and $3p_{5/2}(\Gamma_7)$ can be clearly identified in the figure. The theoretical calculations on the binding energies listed by Molva \textit{et al.}, were performed within the framework of Baldereschi and Lipari theory~\cite{Baldereschi}. As in the case of ZnTe, the ionization energy $E_I$ can be deduced by adding the binding energy of $2p_{5/2}(\Gamma_8)$ to the observed $1s_{3/2}(\Gamma_8) \rightarrow 2p_{5/2}(\Gamma_8)$ transition energy. The ionization energies of Cu and Ag acceptors in CdTe calculated in this way to be 146.25 and 107.85 meV, respectively. The binding energies for the final states of the observed transitions can be calculated by subtracting the experimental transition energies from their ionization energies. The binding energies derived in this way along with those reported by Molva \textit{et al.}~\cite{Molva} are listed in Table~\ref{Tab:CdTe2}.

\section{Fano interactions}
As noted in Section~\ref{AcceptorsExp}, the Lyman lines of group IB acceptors in ZnTe and CdTe, in combination with a simultaneous excitation of the LO phonon and its overtones, overlap the photoionization continuum corresponding to $\hbar \omega \geq E_I$. The line shape of each line in a given phonon replica is characterized by resonance and antiresonance features. The phenomenon was originally recognized and formulated by Breit and Wigner~\cite{Breit} and by Fano~\cite{Fano}, in the context of nuclear transitions and atomic spectroscopy, respectively. Since its discovery, a large number of such situations in diverse physical systems, including semiconductors, have been reported in the literature, e.g., in the Lyman lines of acceptors in Si~\cite{Watkins2}, of Zn$^{-}$ acceptors in Ge~\cite{Piao}, and of 3d-transition-metal-ions in the III-V semiconductors~\cite{Tarhan}.

\begin{figure}
\resizebox{3.4in}{!}{\includegraphics{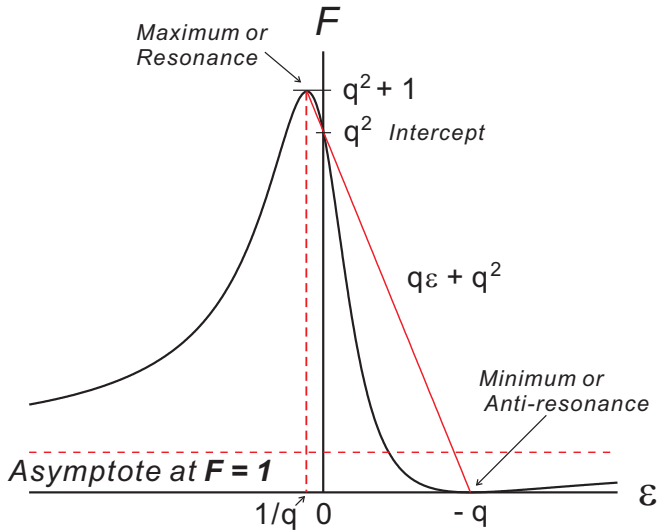}}
\caption{\label{FIG8}Diagram of the Fano ratio, $F(\epsilon, q) = (q + \epsilon)^2 / (1 + \epsilon^2)$, for $q < 0$ with three characteristic features labeled (maximum or resonance, minimum or anti-resonance, and the intercept). The straight line joining the maximum and the minimum of the curve satisfies the equation $F = q \epsilon + q^2$.}
\end{figure}

The detailed description and analysis of the Fano interaction following Piao \textit{et al.}~\cite{Piao} and Tarhan \textit{et al.}~\cite{Tarhan} are given below. The ratio of the Fano absorption to the photoionization absorption is given by the function
\begin{equation}
F(\epsilon, q) = \frac{(q + \epsilon)^2}{(1 + \epsilon^2)},
\end{equation} 
Where the dimensionless, reduced energy variable, $\epsilon$, is defined as
\begin{equation}
\epsilon \equiv \frac{h\nu - [E_{bh} + h\nu_{LO} + F(h\nu)]}{\frac{1}{2} \Gamma},
\end{equation}
$h\nu$ being the photon energy, $E_{bh}$, the corresponding Lyman transition energy of the bound hole, $h\nu_{LO}$, the energy of the zone center longitudinal optical phonon, $F(h\nu)$, the measure of the strength of the coupling between the discrete state and the continuum valence band states, and $\Gamma$, the spectral width of the compound Fano resonant state. The parameter $q$ depends on $F(h\nu)$ and the matrix element between the initial state of the system and the compound Fano state, thus it differs from one bound hole state to another. Following Fig. 3 in Piao \textit{et al.}~\cite{Piao}, reproduced in Fig.~\ref{FIG8}, one can correlate Fano's general formula with the infrared spectra in terms of their three characteristic features: the maximum or resonance ($h\nu_{max}$), the minimum or anti-resonance ($h\nu_{min}$), and the intercept which joining the maximum and the minimum ($h\nu_0$). The values of $F(h\nu)$, $q$, and $\Gamma$ can be expressed as following:
\begin{equation}
F(h\nu) = h\nu_0 - E_{bh} - h\nu_{LO},
\label{Parameter-F}
\end{equation}
\begin{equation}
q = \pm \sqrt{\frac{h\nu_0 - h\nu_{min}}{h\nu_{max} - h\nu_0}},
\label{Parameter-q}
\end{equation}
and
\begin{equation}
\begin{split}
\Gamma &= \frac{2(h\nu_{max} - h\nu_{min})}{q + q^{-1}} \\
&= \left| \frac{2(h\nu_{max} - h\nu_{min})}{\sqrt{\frac{h\nu_0 - h\nu_{min}}{h\nu_{max} - h\nu_0}} + \sqrt{\frac{h\nu_{max} - h\nu_0}{h\nu_0 - h\nu_{min}}}} \right|.
\end{split}
\label{Parameter-Gamma}
\end{equation}
In Eq.~\ref{Parameter-q}, $q$ is positive, if $\nu_{max} > \nu_{min}$ and negative, if $\nu_{max} < \nu_{min}$.

\begin{figure}
\resizebox{3.4in}{!}{\includegraphics{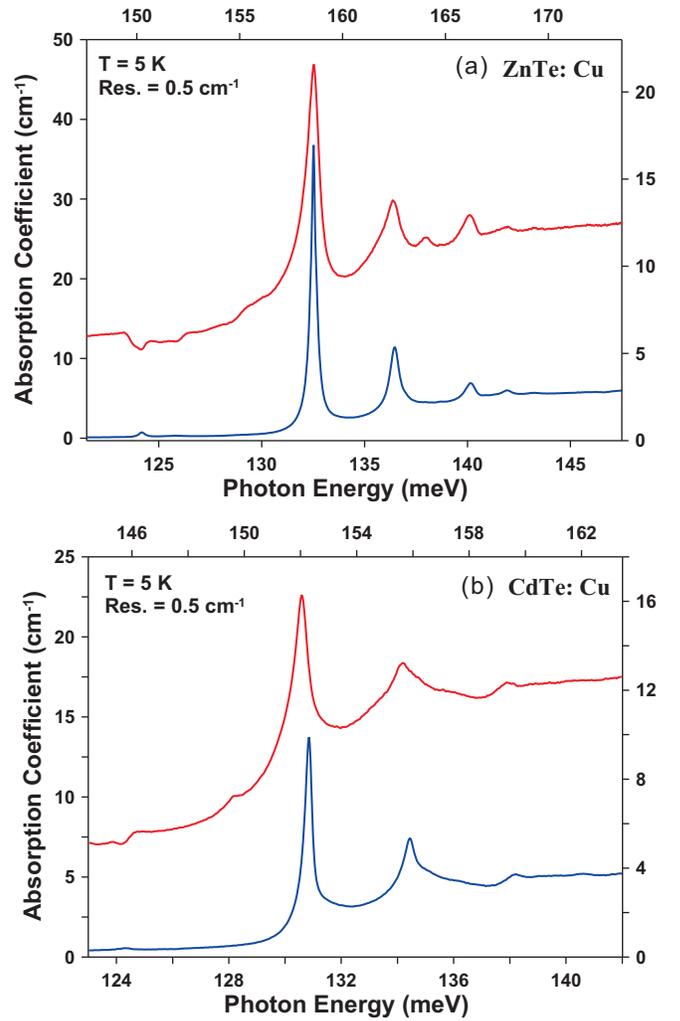}}
\caption{\label{FIG9}The Lyman spectra of (a) ZnTe:Cu and (b) CdTe:Cu and their first LO-phonon assisted overtone compared with the $2p_{3/2}(\Gamma_{8})$ transition in each brought into coincidence. The axes at the bottom and left apply to the Lyman series (lower spectra) while those at the top and right apply to the 1 LO-phonon assisted Fano series (upper spectra).}
\end{figure}

In Fig.~\ref{FIG9}, the Lyman spectra of ZnTe:Cu and CdTe:Cu and the comparison with their first LO-phonon assisted overtone are given. The three characteristic features of each Fano components can be identified in the figure. With these experimental values, one can calculate the corresponding Fano parameters using the equations above. The parameters are tabulated in Table~\ref{Tab:FanoZnTeCu} -~\ref{Tab:FanoCdTeAg} for the 1 LO and 2 LO replicas of Cu and Ag acceptors in ZnTe and CdTe. It is worthy noting while the parameters $q$ and $\Gamma$ are comparable for ZnTe and CdTe, the $F(h\nu)$ values are quite different for ZnTe compared to those of CdTe. 

\begin{table}
\caption{\label{Tab:FanoZnTeCu}Parameters for the 1 LO and 2 LO Fano replicas of the first four Lyman transitions of Cu acceptors in ZnTe, all units being in meV.}
\begin{ruledtabular}
\begin{tabular}{c|cccccccc} 
              &Fano replicas & $F(h \nu)$ & q & $\Gamma$ \\ 
\hline
1 LO replica &$2p_{3/2}(\Gamma_8)^F$     & 0.145 & 0.70   & 0.473 \\
             &$2p_{5/2}(\Gamma_8)^F$     & 0.094 & -5.311 & 0.544\\
             &$2p_{5/2}(\Gamma_7)^F$     & 0.131 & -2.238 & 0.821 \\
             &$3p_{5/2}(\Gamma_8)^F$     & 0.197 & -1.463 & 0.611 \\
\hline
2 LO replica &$2p_{3/2}(\Gamma_8)^{FF}$  & 0.303 & 1.776  & 0.576 \\
             &$2p_{5/2}(\Gamma_8)^{FF}$  & 0.036 & -5.410 & 0.448 \\
             &$2p_{5/2}(\Gamma_7)^{FF}$  & 0.071 & -1.959 & 0.813 \\
             &$3p_{5/2}(\Gamma_8)^{FF}$  & 0.223 & -1.223 & 0.731 \\
\end{tabular}
\end{ruledtabular}
\end{table}

\begin{table}
\caption{\label{Tab:FanoZnTeAg}Parameters for the 1 LO and 2 LO Fano replicas of the first four Lyman transitions of Ag acceptors in ZnTe, all units being in meV.}
\begin{ruledtabular}
\begin{tabular}{c|cccccccc} 
              &Fano replicas & $F(h \nu)$ & q & $\Gamma$ \\ 
\hline
1 LO replica &$2p_{3/2}(\Gamma_8)^F$     & 0.311 & 1.835  & 0.459 \\
             &$2p_{5/2}(\Gamma_8)^F$     & 0.590 & -6.546 & 0.412\\
             &$2p_{5/2}(\Gamma_7)^F$     & 0.636 & -6.126 & 0.548 \\
             &$3p_{5/2}(\Gamma_8)^F$     & 0.764 & -3.890 & 0.271 \\
\hline
2 LO replica &$2p_{3/2}(\Gamma_8)^{FF}$  & 0.541 & 0.703  & 0.428 \\
             &$2p_{5/2}(\Gamma_8)^{FF}$  & 0.652 & -5.204 & 0.535 \\
             &$2p_{5/2}(\Gamma_7)^{FF}$  & 0.681 & -3.250 & 1.022 \\
             &$3p_{5/2}(\Gamma_8)^{FF}$  & 0.874 & -2.245 & 0.712 \\
\end{tabular}
\end{ruledtabular}
\end{table}

\begin{table}
\caption{\label{Tab:FanoCdTeCu}Parameters for the 1 LO and 2 LO Fano replicas of the first four Lyman transitions of Cu acceptors in CdTe, all units being in meV.}
\begin{ruledtabular}
\begin{tabular}{c|cccccccc} 
              &Fano replicas & $F(h \nu)$ & q & $\Gamma$ \\ 
\hline
1 LO replica &$2p_{3/2}(\Gamma_8)^F$     & 0.120  & 0.100  & 0.620 \\
             &$2p_{5/2}(\Gamma_8)^F$     & -0.208 & -5.305 & 0.504 \\
             &$2p_{5/2}(\Gamma_7)^F$     & -0.189 & -6.568 & 0.783 \\
             &$3p_{5/2}(\Gamma_8)^F$     & -0.194 & -1.800 & 0.382 \\
\hline
2 LO replica &$2p_{3/2}(\Gamma_8)^{FF}$  & 0.829  & 2.884  & 0.878 \\
             &$2p_{5/2}(\Gamma_8)^{FF}$  & -0.426 & -4.960 & 0.473 \\
             &$2p_{5/2}(\Gamma_7)^{FF}$  & -0.037 & -2.511 & 1.714 \\
             &$3p_{5/2}(\Gamma_8)^{FF}$  & -0.253 & -1.652 & 0.966 \\
\end{tabular}
\end{ruledtabular}
\end{table}

\begin{table}
\caption{\label{Tab:FanoCdTeAg}Parameters for the 1 LO replicas of the first three Lyman transitions of Ag acceptors in CdTe, all units being in meV.}
\begin{ruledtabular}
\begin{tabular}{c|cccccccc} 
              &Fano replicas & $F(h \nu)$ & q & $\Gamma$ \\ 
\hline
1 LO replica &$2p_{5/2}(\Gamma_8)^F$     & -0.282 & -3.193 & 0.866 \\
             &$2p_{5/2}(\Gamma_7)^F$     & -0.061 & -2.649 & 1.190 \\
             &$3p_{5/2}(\Gamma_8)^F$     & -0.207 & -1.490 & 0.469 \\
\end{tabular}
\end{ruledtabular}
\end{table}

\section{Concluding Remarks}

The present study is focused on the Lyman spectra of group IB acceptors (Cu, Ag, and Au), substitutionally incorporated in ZnTe and CdTe by replacing the cations. The sharpness of the Lyman transitions observed at the cryogenic temperatures indicates that wavefunctions of the acceptor-bound holes do not overlap. The Cu and Ag acceptors in both ZnTe and CdTe yielded sharp and intense transitions terminating at excited states preceding the photoionization threshold. The onset of the photoionization is in excellent agreement with the sum of the transition energy and the binding energy of its final state calculated in the EMT~\cite{Kohn}. 

As in the cases of chalcogens in Si~\cite{Olajos,Kleverman} and 3d transition metal ions in III-V semiconductors~\cite{Tarhan}, the group IB impurities enter II-VI semiconductors as deep defect centers. The EMT can be used to determine the shallow states of these defects ($p$-like states) where the electron is bound to the defect by a screened Coulomb potential, which give rise to a hydrogen-like energy spectrum. As can be seen in Tables~\ref{Tab:ZnTe2} and ~\ref{Tab:CdTe2}, the binding energies for the same $p$-like states of Cu, Ag, and Au are all very close to their theoretical values. However, the EMT breaks down when the electron wave function has a non-negligible amplitude ($s$-like states) in the central cell region where it is affected by the short-range impurity potential, not accounted for in EMT, and the difference between the experimental and theoretical values can be more than four times (e.g., in case of Au acceptors in ZnTe).    

The Fano resonance interaction in semiconductors is one of the most interesting example of electron-phonon interactions~\cite{Klein}. The strong ionicity of the II-VI semiconductors clearly favors this type of interaction; up to fourth overtone of the Fano replicas were observed in both ZnTe and CdTe as compared to three overtones in III-Vs~\cite{Tarhan} and one in Si and Ge~\cite{Watkins2,Piao}. The strength of the coupling between the discrete state and the photoionization continuum as reflected in the parameter $F(h\nu)$ is also considerably larger for the II-VIs than those for the III-Vs and Si and Ge. 

The group V elements enter the II-VI semiconductors by replacing the ions and act as acceptors, for example, As replaces Te in ZnTe~\cite{Saminadayar}. It is of great interest to explore the Lyman series of these acceptors as well as their Fano replicas. In their study, infrared absorption spectroscopy along with Raman spectroscopy and photoluminescence would provide a more complete and accurate picture for the electronic and optical properties of these defects.

\end{document}